\documentclass[preprint]{revtex4}
\usepackage{graphicx}
\newcommand{\be}{\begin{equation}}
\newcommand{\ee}{\end{equation}}

\begin{document}

\title{Localized modes on an Ablowitz-Ladik nonlinear impurity}

\author{Mario I. Molina}

\affiliation{Departmento de F\'{\i}sica, Facultad de Ciencias,
Universidad de Chile, Santiago, Chile}

\begin{abstract}
We study localized modes on a single Ablowitz-Ladik impurity
embedded in the bulk or at the surface of a one-dimensional linear
lattice. Exact expressions are obtained for the bound state
profile and energy. Dynamical excitation of the localized mode
reveals exponentially-high amplitude oscillations of the spatial profile at
the impurity location. The presence of a surface increases the
minimum nonlinearity to effect a dynamical selftrapping.
\end{abstract}

%\ocis{060.4370;190.4350;190;6135}
%PACS: 73.20.Hb, 03.65.Ge, 71.55.2i
\maketitle

\newpage

The study of nonlinear dynamics in discrete systems has attracted
a special attention recently due to novel physics and possible
interesting applications~\cite{review}. Among these systems, we find
the integrable discretized version of the continuum NLS
equation, the so-called Ablowitz-Ladik (AL) equation\cite{al}:
\be
i {d C_{n}\over{dt}} + ( V + \mu |C_{n}|^{2} )( C_{n+1} + C_{n-1} ) = 0
\label{eq:0}
\ee
This integrable version support moving, nonlinear, spatially-localized 
excitations in the form of lattice solitons, found through the use
of the inverse scattering transform method. The AL equations constitute  
a starting point for many studies on the interplay of disorder,  
nonlinearity and discreteness. For instance, when examining the effects of disorder,
a well-known approach is to assume a perturbative approach and try
to compute the evolution of the soliton parameters\cite{garnier}. When the scale of the
disorder is high, this approach is no longer tenable and one must resort to 
numerical schemes. On the other hand, when nonlinearity is large, the spatial
soliton profile is well localized in space, meaning that only a small number of sites
around the soliton center are effectively nonlinear. The system then looks very
similar to a linear system  containing a small cluster of nonlinear sites, or even
a single nonlinear impurity. This simplified system is now amenable to exact 
mathematical treatment, and the influence of other potentially competing effects, such as
dimensionality, boundary effects, noise, etc., can be more
easily studied without losing the essential physics. This approach has been
successfully used for the DNLS equation\cite{molina_dnls_impurity},
\be
i {d C_{n}\over{dt}} + V( C_{n+1} + C_{n-1})  + \gamma |C_{n}|^{2} C_{n}= 0,
\label{eq:00}
\ee
 where
it was predicted that, for a semi-infinite nonlinear chain, the presence of a surface
would increase the amount of nonlinearity required to form a localized surface mode. 
This was subsequently observed in later studies\cite{makris1,vmk_semi}. When used
for the two-dimensional semi-infinite square lattice, this procedure predicted that
this time, the presence of a boundary would decrease the minimum nonlinearity needed to
create a surface localized mode\cite{molina_semi_2D}. This was later found to be the case\cite{kevrekidis}.

In this Letter, we introduce a novel type of nonlinear defect in a
one-dimensional discrete chain, this time using the framework of the AL
equation (\ref{eq:0}).

We consider a one-dimensional array of linear sites, containing a
single, Ablowitz-Ladik impurity located at site $n_{0}$. In the
tight-binding framework, the evolution equation for the amplitude
is given by
\be
i {d C_{n}\over{dt}} + ( V + \delta_{n,n_{0}} \mu |C_{n}|^{2} )( C_{n+1} + C_{n-1} ) = 0
\label{eq:1}
\ee
where $C_{n}$ is the complex amplitude at site $n$, $V$ is the
nearest-neighbor coupling coefficient, and $\mu$ is the
Ablowitz-Ladik (AL) parameter. We will be interested in
stationary-state solutions of the form $C_{n}(t) = C_{n} \exp(i
\omega t)$. This leads to the system of equations: \be -\omega
C_{n} + ( V + \delta_{n,n_{0}} \mu |C_{n}|^{2} )( C_{n+1} +
C_{n-1} ) = 0. \label{eq:2} \ee

From Eq.(\ref{eq:1}) it can be easily proven that the norm
\be
 {\cal N} = (V/\mu) \log( 1 + (\mu/V) |C_{0}|^{2} ) +
{\sum_{n}}^{'} |C_{n}|^{2},
\ee
is a conserved quantity, where the prime in the sum indicates that
the sum is carried out over all sites, excepting the impurity
site, $n=n_{0}$. We normalize the time to $\tau = V t$ and the
probability amplitude to $\phi_{n} = C_{n} /\sqrt{{\cal N}}$. With
these definitions, Eq.(\ref{eq:1}) simplifies to
\be
i {d \phi_{n}\over{d\tau}} + ( 1 + \delta_{n,n_{0}} \nu
|\phi_{n}|^{2} )( \phi_{n+1} + \phi_{n-1} ) = 0 \label{eq:new1}
\ee %
where $\nu \equiv {\cal N}\mu/V$. The normalization condition
becomes
\be
1 = (1/\nu) \log(1 + \nu |\phi_{n_{0}}|^{2}) +
{\sum_{n}}^{'} |\phi_{n}|^{2}.\label{eq:3}
\ee
The equation for the stationary state, acquires now its dimensionless form:
\be
-\beta\ \phi_{n} + ( 1 + \delta_{n,n_{0}} \nu |\phi_{n}|^{2} )( \phi_{n+1} + \phi_{n-1} ) = 0,
\label{eq:4}
\ee
where, $\beta \equiv \omega/V$.

We will focus on two special cases, (i) Impurity in the ``bulk'' and (ii) ``surface'' impurity.

{\bf Impurity in the ``bulk''}: In this case, $-\infty < n
< \infty$ and without loss of generality, we choose $n_{0}=0$. 
We pose a solution in the form $\phi_{n}
= A\ \xi^{|n|}$, where $0 < |\xi| < 1$. After inserting this
ansatz into Eq.(\ref{eq:4}), one obtains $\beta = 2 \xi (1 + \nu
A^{2})$ and $\beta = \xi + (1/\xi)$. After solving for $\xi$, one
obtains
\be
\xi^{2} = {1\over{1 + 2 \nu A^{2}}}
\ee
On the other hand, from the normalization condition, Eq.(\ref{eq:3}), one
obtains the relation
\be
1 = {1\over{\nu}}\log(1 + \nu A^{2}) + {2 A^{2} \xi^{2}\over{1-\xi^{2}}}.
\ee
After combining these last two equations, one obtains $\xi = \pm [2 \exp(\nu-1) -1
]^{-1/2}$, and $A = ((\exp(\nu-1)-1)/\nu)^{1/2}$, which implies
\be
\phi_{n} = (\pm 1)^{n}\ \left(
{\exp(\nu-1)-1 \over{\nu}}\right)^{1/2}\ \left( 2 \exp(\nu-1) -1
\right)^{-|n|/2}. \label{eq:phin}
\ee
The dimensionless bound
state energy is
\be
\beta = \pm \left( [2 \exp(\nu-1) -1 ]^{-1/2} + [2
\exp(\nu-1) -1 ]^{1/2} \right).\label{eq:beta}
\ee
As can be seen from Eq.(\ref{eq:phin}), a localized bound state is
possible provided $\nu > \nu_{c}=1$, and for a given $\nu$, there
is an {\em unstaggered} ({\em staggered}) version of the bound
state for $\beta>2\ (<-2)$. Fig.1 shows a couple of profiles
$\phi_{n}$ and their staggered versions, for two different
dimensionless nonlinearity parameters $\nu$. In Fig.2 we show $\xi$ and the
bound state energy as a function of nonlinearity. Standard linear stability analysis reveals
that this stationary localized state is stable.
%
%%%%% FIG.1 %%%%%%%%%%%
\begin{figure}[t]
\includegraphics[scale=0.75]{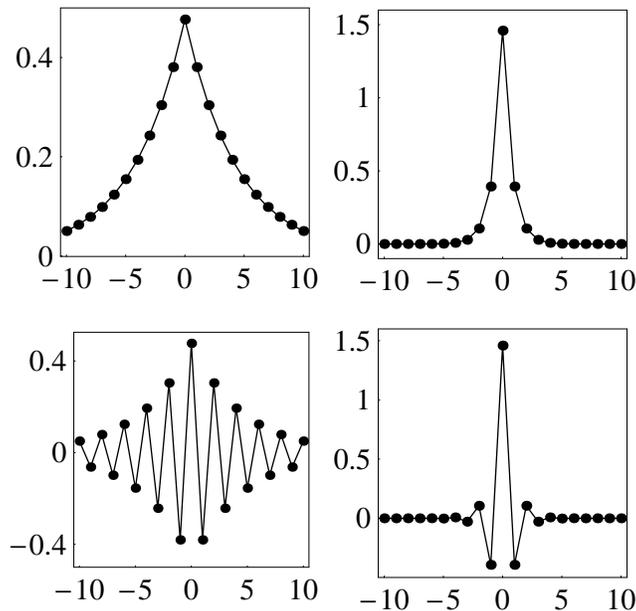}
\caption{Impurity in bulk: localized modes for $\nu=1.25$ (left
column) and $\nu=1.5$ (right column). The top (bottom) row shows
the unstaggered (staggered) versions of the mode.} \label{fig1}
\end{figure}
%%%%%%%%%%%%%%%%%%%%%%

An interesting feature arises when we consider the dynamical
excitation of a localized state. In this case, one considers
Eq.(\ref{eq:new1}) for a highly localized initial condition,
chosen as $\phi_{n}(0) = \delta_{n,0} \sqrt{(\exp(\nu)-1)/\nu}$.
This choice corresponds to the one that saturates the normalization condition,
Eq.(\ref{eq:3}).  
Examination of the ensuing dynamics reveals that at low
nonlinearity values, the excitation tends to diffract across the
array, while for higher nonlinearities, it tends to selftrap at
the impurity site, with a high-amplitude oscillation, as Fig.3
clearly shows. The magnitude and frequency of these oscillations
increase as the nonlinearity parameter $\nu$ is increased. We have
checked numerically the persistence of this breathing phenomenon
for long times, and believe that it can be understood from the
special form of the normalization condition, Eq.(\ref{eq:3}): A
small change in the sum of the square amplitudes at sites other 
than the impurity site will
bring about a large change of the amplitude at the impurity site,
due to the logarithmic dependency of the latter. To be more
precise, let us assume that shortly after launching the initial 
excitation, a certain amount of radiation is emitted causing 
${\sum}' |\phi_{n}|^{2}\rightarrow {\sum}' |\phi_{n}|^{2} - \Delta$; 
then it can be easily proven that $|\phi_{0}|^{2}\rightarrow |\phi_{0}|^{2} + (1/\nu) (\exp(\nu
\Delta) -1)$. Thus, it is the particular form of the AL
nonlinearity that amplifies the breathing oscillations
exponentially at the impurity site. We have also computed the
long-time average probability at the initial site, as a function
of nonlinearity strength. For our relatively short chain (100
sites), there is no sharp selftrapping threshold, although there
is an inflexion point around $\nu=7$, as Fig.3 shows.
%%%%% FIG.2 %%%%%%%%%%%
\begin{figure}[t]
\includegraphics[scale=0.8]{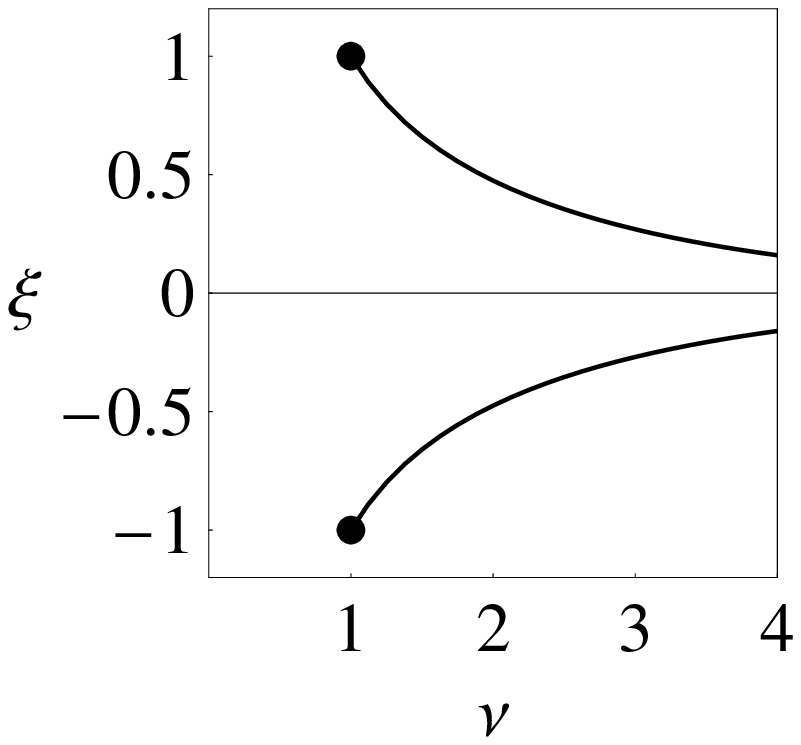}
\includegraphics[scale=0.8]{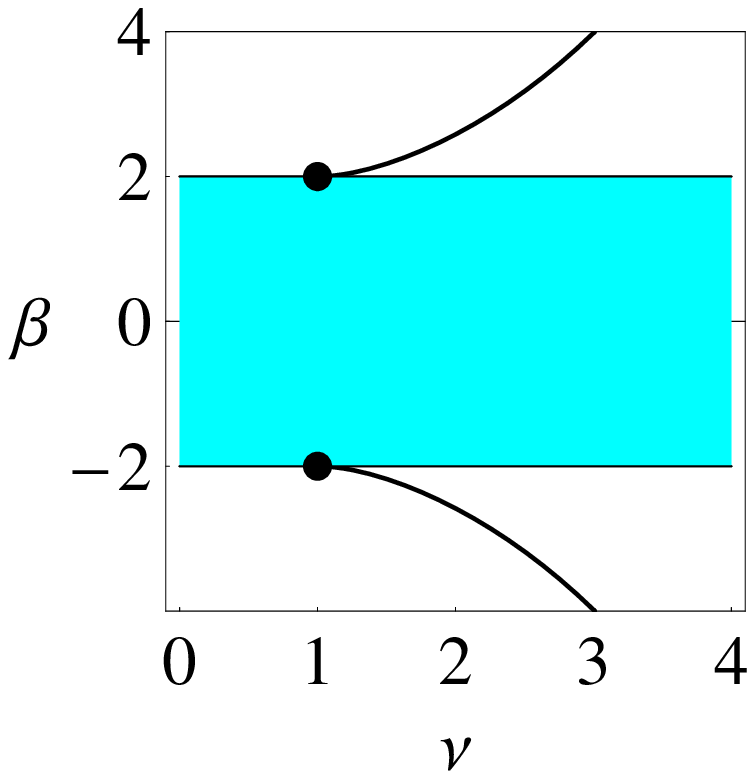}
\caption{Impurity in bulk. Left: $\xi$ as a function of $\nu$ for localized mode.
Right: Bound state energy of localized mode as
a function of nonlinearity parameter. The shaded area marks the position of the 
linear band, while the upper (lower) curve corresponds to the unstaggered (staggered)
mode. The black dot marks the
position of $\nu_{c}=1$.} \label{fig2}
\end{figure}
%%%%%%%%%%%%%%%%%%%%%%

%%%% FIG.3 %%%%%%%%%%%
\begin{figure}[t]
\includegraphics[scale=0.5]{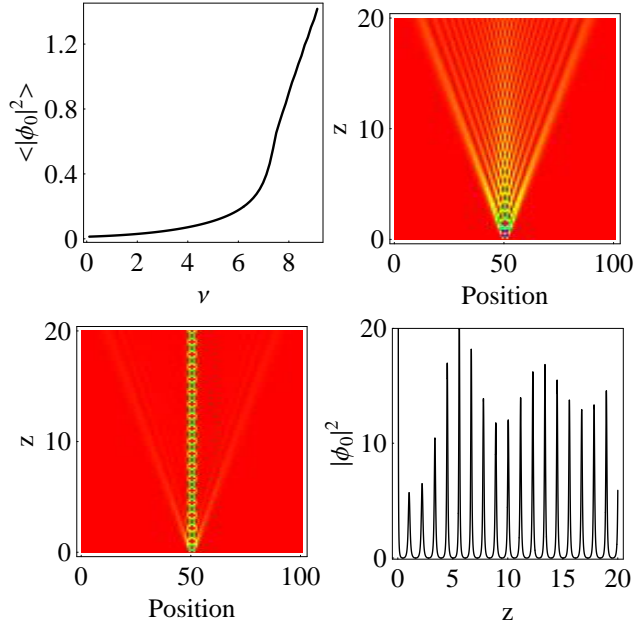}
\caption{Impurity in bulk. Top left: Long-time average probability
at impurity site. Top right: Evolution of initial localized
excitation across the lattice for $\nu=2$. Bottom left: Evolution
for $\nu=8$. Bottom right: Evolution of amplitude at impurity site
for $\nu=8$. } \label{fig3}
\end{figure}
%%%%%%%%%%%%%%%%%%%%%%

{\bf Surface impurity}: We now consider the case when the impurity
is at the very beginning of a semi-infinite lattice. We relabel
the previous chain, so that the first site is now at $n_{0}=0$. The
dimensionless stationary-state equations read now
\be -\beta\ \phi_{0} + ( 1 + \nu |\phi_{0}|^{2}) \phi_{1} =
0\label{eq:4b} \ee
\be -\beta\ \phi_{n} + ( \phi_{n+1} + \phi_{n-1} ) = 0,\hspace{1cm}n=0,1,2,\ldots
\label{eq:5}
\ee

We proceed as before and pose a solution of the form $\phi_{n} = A
\xi^{n}$, where $0<|\xi|<1$ and $n=0,1,2,\dots$. After replacing
this ansatz into Eq.(\ref{eq:4b}) and (\ref{eq:5}), one obtains
$\beta=(1+\nu A^{2}) \xi$ and $\beta=\xi+(1/\xi)$, which implies
\be%
\xi^{2} = {1\over{\nu A^{2}}}\label{eq:6}
\ee
On the other hand, from the normalization condition,
Eq.(\ref{eq:3}), we have
\be
1 = {1\over{\nu}}\log(1 + \nu A^2) + {A^{2}
\xi^{2}\over{1-\xi^{2}}}\label{eq:7}
\ee
From Eqs. (\ref{eq:6}) and (\ref{eq:7}), we obtain a transcendental
equation for $\xi$:
\be
\nu = \log\left(1 + {1\over{\xi^{2}}}\right) +
{1\over{1-\xi^{2}}}\label{eq:8}
\ee
Simple analysis shows that there is a critical nonlinearity value  
$\nu_{c}=(3/2)+\log(4)\approx 2.9$, such that, for $\nu<\nu_{c}$ there
is no bound state, at $\nu=\nu_{c}$ there is exactly one bound state, while
for $\nu>\nu_{c}$ there are two bound states. One of these states, becomes more
narrow and its energy detaches from the linear band as nonlinearity is increased, while
the second one becomes wider and its energy approaches the linear band upon increase in 
nonlinearity (see Fig.5 below). Straightforward linear stability analysis reveals that the former 
state is stable, while the latter is unstable.

The bound state mode is given by
\be
\phi_{n} = {1\over{\sqrt{\nu}}}
\xi(\nu)^{n-1}\hspace{1cm}n=0,1,\dots
\ee
where $\xi$ has to be found numerically from Eq.(\ref{eq:8}), for a given $\nu>\nu_{c}$. 
It is possible, however, to derive a very simple, yet accurate,
approximation for $\xi=\xi(\nu)$, as follows: We start from Eq.(\ref{eq:8})
re-written as
\be
\exp(\nu) = \left( {1+\xi^{2}\over{\xi^{2}}} \right) \exp(1/(1-\xi^{2}))\label{eq:88}
\ee
Now, since $0<|\xi|<1$,
it makes sense to expand around $\xi=0$. To
fourth-order in $\xi$, Eq.(\ref{eq:88}) becomes
\be
\xi^{2} e^{\nu-1} \approx 1 + 2\ \xi^{2} + (5/2)\ \xi^{4}
\ee
which implies,
%%%% FIG.4 %%%%%%%%%%%
\begin{figure}[b]
\includegraphics[scale=0.8]{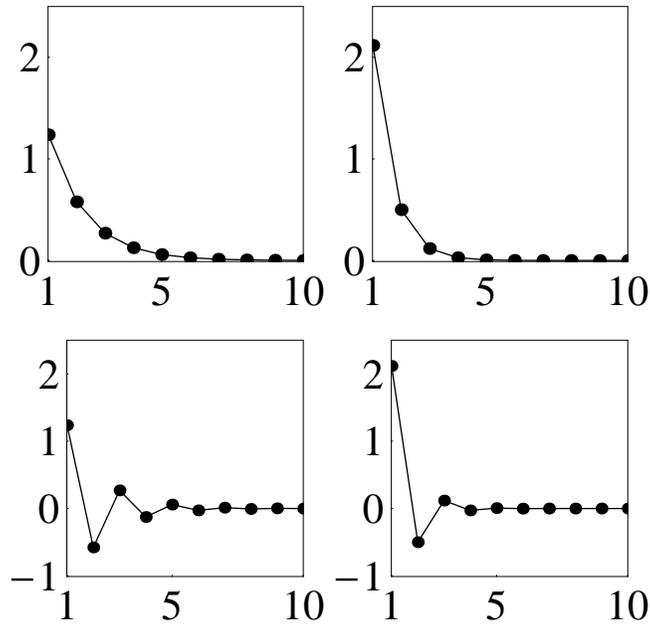}
\caption{Surface impurity: localized modes for $\nu=3$ (left
column) and $\nu=4$ (right column). The top (bottom) row shows
the unstaggered (staggered) versions of the mode. } \label{fig4}
\end{figure}
%%%%%%%%%%%%%%%%%%%%%%
%
\be
\xi(\nu) \approx \pm \left( {1\over{5}} ( \exp[\nu-1]-2-( \exp[2(\nu-1)]-4 \exp[\nu-1]-6 )^{1/2} )\right)^{1/2}
\label{eq:approx}
\ee

Numerical comparison with the exact value, reveals that the relative
percentage error of approximation (\ref{eq:approx}) is less than
$3\%$ for $\nu>3$.

Figure 4 shows some amplitude profiles in the vicinity of the lattice surface
for a couple of different $\nu$ values. Figure 5 shows the numerical solution for
$\xi$ and the localized state energy as a function of nonlinearity. As before,
values of $\beta$ above (below) the band give rise to unstaggered
(staggered) states.

Comparison between Figs. 
2 and 5 reveals that, as far as stationary localized modes is 
concerned, the presence of a surface increases the minimum amount of nonlinearity needed to
create a bound state. The boundary is acting as a repulsive
surface, similar to what has been observed earlier in
semi-infinite DNLS systems\cite{molina_dnls_impurity,vmk_semi}

%%%% FIG.5 %%%%%%%%%%%
\begin{figure}[t]
\includegraphics[scale=0.75]{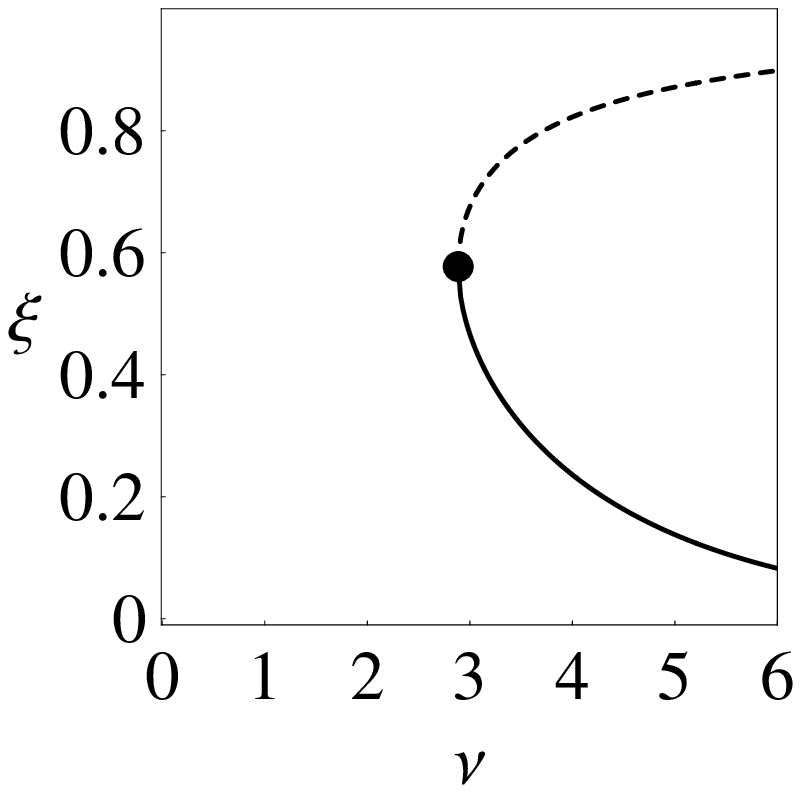}
\includegraphics[scale=0.75]{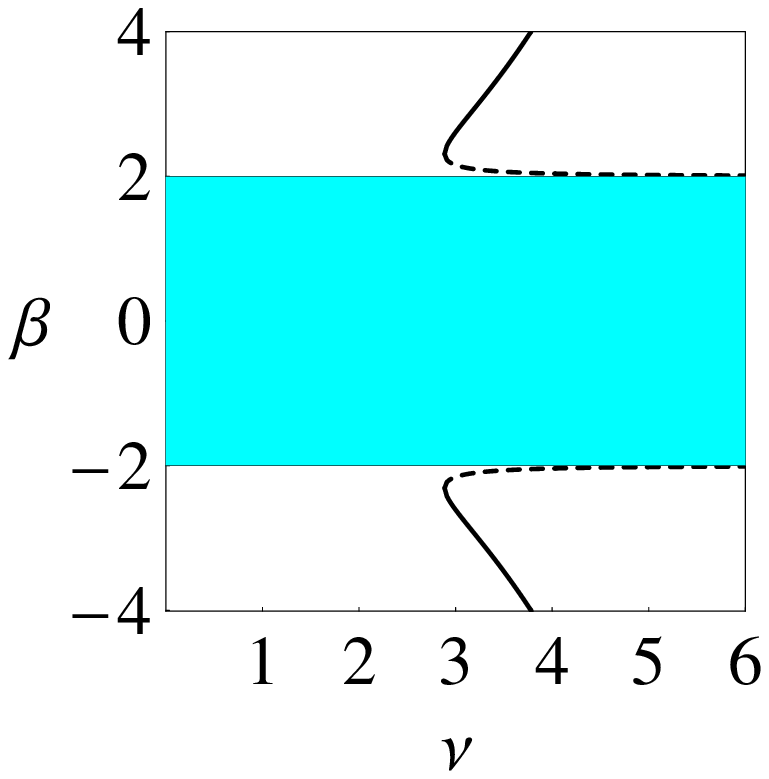}
\caption{Surface impurity. Left: Numerical solution
for $\xi$ in terms of $\nu$. 
Right: Bound state energies of localized modes as
a function of nonlinearity parameter. Solid(dashed) curve denotes
stable(unstable) solution.} \label{fig5}
\end{figure}
%%%%%%%%%%%%%%%%%%%%%%

Finally, we examine the dynamics of an excitation initially
localized at the surface of the system $n=0$. The idea is to
determine how the presence of a boundary affects the dynamical
creation of a surface localized mode. As before, we take
$\phi_{n}(0)=\sqrt{(\exp(\nu)-1)/\nu}\ \delta_{n,0}$ and examine
the average probability remaining at the initial site for long
times, as well as the behavior of the amplitude at the impurity.
Results are displayed in Fig.6, which is qualitatively similar to
its bulk counterpart, Fig.3.
%%%% FIG.6 %%%%%%%%%%%
\begin{figure}[t]
\includegraphics[scale=0.5]{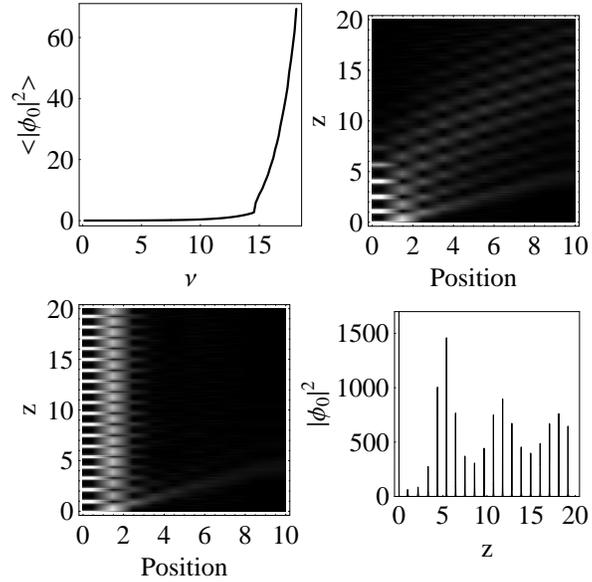}
\caption{Surface impurity. Top left: Long-time average probability
at impurity site. Top right: Evolution of initial localized
excitation across the lattice for $\nu=14$. Bottom left: Evolution
for $\nu=15$. Bottom right: Evolution of amplitude at impurity
site for $\nu=15$. } \label{fig6}
\end{figure}
%%%%%%%%%%%%%%%%%%%%%%
As before,  we observe diffraction behavior for small nonlinearity
values and selftrapping at large $\nu$ values (at approximately
$\nu\sim 14.6$). In the last case, we also observe large-amplitude
oscillations at the impurity site. The main difference with the
bulk case, is that we need now substantially larger $\nu$ values
to effect selftrapping.

In conclusion, we have examined the stationary-state and dynamical 
localized modes residing on a AL-like impurity, embedded well inside 
and at the surface of a one-dimensional discrete lattice. For both cases, the dynamical
localized mode displays high-amplitude (exponential) oscillations at the impurity site, due to
the particularly asymmetric form of the coupling between the impurity and its neighbors.
The presence of a surface, on the other hand, increases the amount of minimum nonlinearity 
needed to create a localized mode, in agreement with previous studies on one-dimensional DNLS systems.

\section{Acknowledgments} 
This work has been supported by Fondecyt grant 1080374 in Chile. The author 
is grateful to M. J. Ablowitz for useful discussions.

\newpage

\section*{\centerline{List of Figure Captions}}

\noindent
Figure 1:\ \ Impurity in bulk: localized modes for $\nu=1.25$ (left
column) and $\nu=1.5$ (right column). The top (bottom) row shows
the unstaggered (staggered) versions of the mode.
\vspace{0.4cm}

\noindent
Figure 2:\ \ (Color online) Impurity in bulk. Left: $\xi$ as a function of $\nu$ for localized mode.
Right: Bound state energy of localized mode as
a function of nonlinearity parameter. The shaded area marks the position of the 
linear band, while the upper (lower) curve corresponds to the unstaggered (staggered)
mode. The black dot marks the
position of $\nu_{c}=1$.
\vspace{0.4cm}

\noindent
Figure 3:\ \ (Color online) Impurity in bulk. Top left: Long-time average probability
at impurity site. Top right: Evolution of initial localized
excitation across the lattice for $\nu=2$. Bottom left: Evolution
for $\nu=8$. Bottom right: Evolution of amplitude at impurity site
for $\nu=8$. 
\vspace{0.4cm}

\noindent
Figure 4:\ \ Surface impurity: localized modes for $\nu=3$ (left
column) and $\nu=4$ (right column). The top (bottom) row shows
the unstaggered (staggered) versions of the mode. 
\vspace{0.4cm}

\noindent
Figure 5:\ \ (Color online) Surface impurity. Left: Numerical solution
for $\xi$ in terms of $\nu$. 
Right: Bound state energies of localized modes as
a function of nonlinearity parameter. Solid(dashed) curve denotes
stable(unstable) solution.
\vspace{0.4cm}

\noindent
Figure 6:\ \ Surface impurity. Top left: Long-time average probability
at impurity site. Top right: Evolution of initial localized
excitation across the lattice for $\nu=14$. Bottom left: Evolution
for $\nu=15$. Bottom right: Evolution of amplitude at impurity
site for $\nu=15$. 


\begin{thebibliography}{99}

\bibitem{review}  See, e.g., D.N. Christodoulides, F. Lederer, and Y.
Silberberg, Nature {\bf 424}, 817 (2003) and references therein.

\bibitem{al}
M. J. Ablowitz and J. F. Ladik, J. Math. Phys. {\bf 16}, 598, (1975).

\bibitem{garnier}
J. Gernier, Phys. Rev. E {\bf 63}, 026608 (2001).

\bibitem{molina_dnls_impurity}
M. I. Molina, Phys. Rev. B {\bf 71}, 035404 (2005).

\bibitem{makris1}
K. G. Makris, S. Suntsov, D. N. Christodoulides and G. I. Stegeman,
Opt. Lett. {bf 30}, 2466 (2005).

\bibitem{vmk_semi}
M. I. Molina, R. A. Vicencio and Y. S. Kivshar, Optt. Lett. {\bf 31}, 1693 (2006).

\bibitem{molina_semi_2D}
M. I. Molina, Phys. Rev. B {\bf 74}, 045412 (2006).

\bibitem{kevrekidis}
H. Susanto, P. G. Kevrekidis, B. A. Malomed, R. Carretero-Gonz\'{a}lez and D. J. Frantzeskakis,
Phys. Rev. E. {\bf 75}, 056605 (2007).

\end{thebibliography}
\end{document}